\documentclass{PoS}
\usepackage{wrapfig}
\newcommand{\beq}{\begin{eqnarray}}
\newcommand{\eeq}{\end{eqnarray}}

\title{Quark number density at imaginary chemical potential and its extrapolation to large real chemical potential by the effective model}

\ShortTitle{Quark number density at imaginary chemical potential...}

\author{\speaker{Junichi Takahashi}\\
        Department of Physics, Graduate School of Sciences, Kyushu University, Fukuoka 812-8581, Japan\\
        E-mail: \email{takahashi@phys.kyushu-u.ac.jp}}

\author{Junpei Sugano\\
        Department of Physics, Graduate School of Sciences, Kyushu University, Fukuoka 812-8581, Japan\\
        E-mail: \email{sugano@phys.kyushu-u.ac.jp}}

\author{Masahiro Ishii\\
        Department of Physics, Graduate School of Sciences, Kyushu University, Fukuoka 812-8581, Japan\\
        E-mail: \email{ishii@phys.kyushu-u.ac.jp}}

\author{Hiroaki Kouno\\
        Department of Physics, Saga University, Saga 840-8502, Japan\\
        E-mail: \email{kounoh@cc.saga-u.ac.jp}}

\author{Masanobu Yahiro\\
        Department of Physics, Graduate School of Sciences, Kyushu University, Fukuoka 812-8581, Japan\\
        E-mail: \email{yahiro@phys.kyushu-u.ac.jp}}

\abstract{We evaluate quark number densities at imaginary chemical potential by lattice QCD with clover-improved two-flavor Wilson fermion. 
The quark number densities are extrapolated to the small real chemical potential region by assuming some function forms. 
The extrapolated quark number densities are consistent with those calculated at real chemical potential with the Taylor expansion method for the reweighting factors. 
In order to study the large real chemical potential region, we use the two-phase model consisting of the quantum hadrodynamics model for the hadron phase and the entanglement-PNJL model for the quark phase. 
The quantum hadrodynamics model is constructed to reproduce nuclear saturation properties, while the entanglement-PNJL model reproduces well lattice QCD data for the order parameters such as the Polyakov loop, the thermodynamic quantities and the screening masses. 
Then, we calculate the mass-radius relation of neutron stars and explore the hadron-quark phase transition with the two-phase model. 
}

\FullConference{The 32nd International Symposium on Lattice Field Theory,\\
		23-28 June, 2014\\
		Columbia University New York, NY}

\begin{document}
\section{Introduction}
QCD phase diagram and the neutron stars (NSs) are important topics for studying the properties of QCD in finite chemical potential ($\mu$) region.
The quark number density is a fundamental quantity in this region and plays an important role of determining the strength of the vector-type interaction in the effective model.
However, lattice QCD (LQCD) as a first-principle calculation has the sign problem at finite $\mu$.
So far, the quark number density has been studied by the Taylor expansion method for the reweighting factor at real $\mu$ with Wilson-type quark actions~\cite{Ejiri} and by the analytic continuation from imaginary to real $\mu$ with staggered-type quark actions~\cite{D'Elia1,D'Elia2}.
\\
\indent
In this study, we calculate the quark number density at imaginary $\mu$ ($\mu_{\mathrm{I}}\equiv i\mu$) with Wilson-type quark action.
Then, the quark number density is extrapolated to real $\mu$ region with analytic continuation by assuming some function forms\footnote{Please see Sec.~\ref{QND} in detail}.
In the imaginary $\mu$ region, QCD has two characteristic 
\begin{wrapfigure}[12]{r}{7.2cm}
\vspace{-0.2cm}
\includegraphics[width=5cm,height=7.2cm,angle=90]{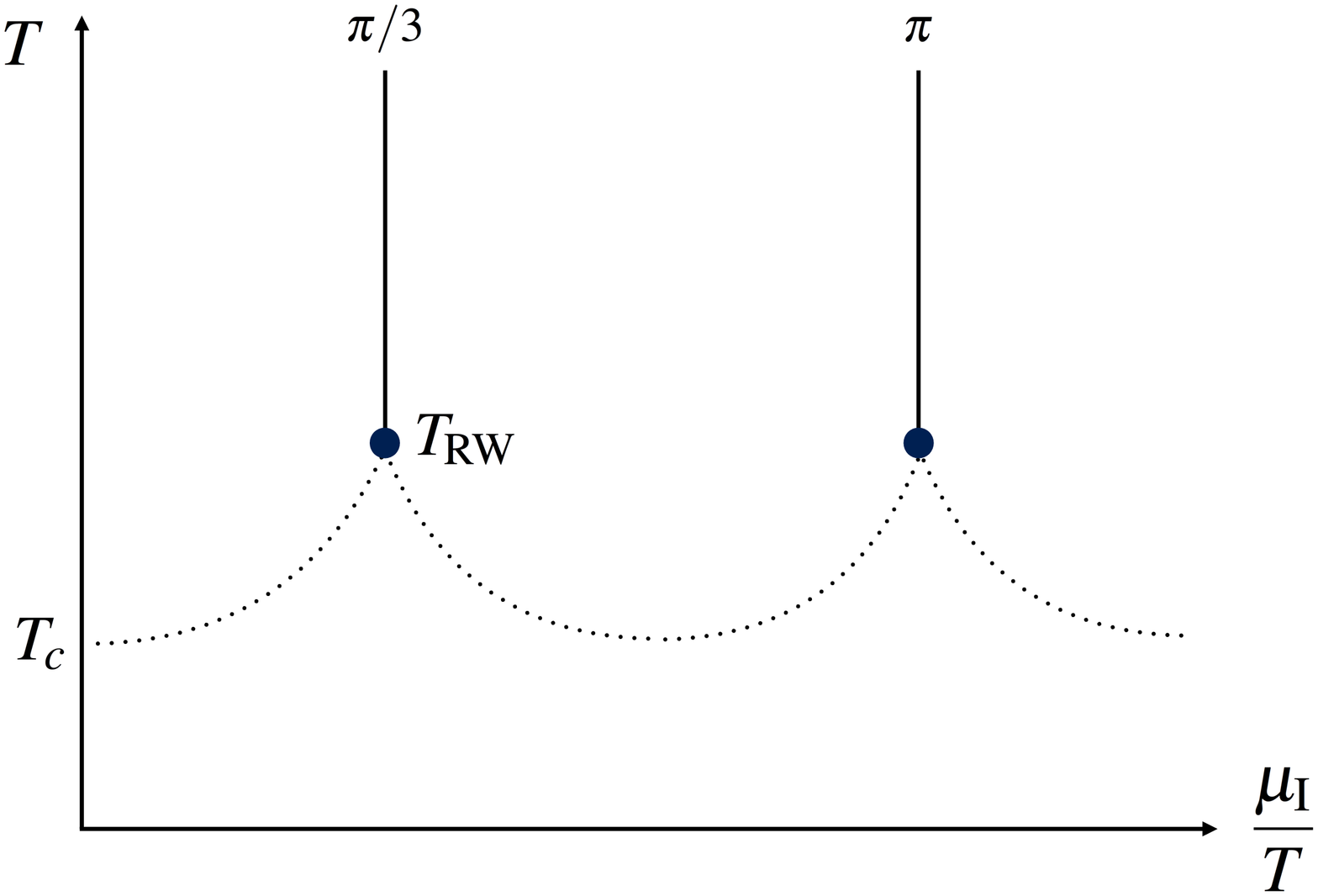}
\vspace{-1.2cm}
\caption{
QCD phase diagram in the imaginary $\mu$ region.
}
\label{fig:mu_I-QCD_phase_diagram}
\end{wrapfigure}
properties: one is the Roberge-Weiss (RW) periodicity in the QCD partition function and the other is so called the RW transition of first order~\cite{RW}.
The partition function has a periodicity of $2\pi/3$ as a function of $\mu_{\mathrm{I}}/T$, namely $Z(\mu_{\mathrm{I}}/T) =Z(\mu_{\mathrm{I}}/T+2\pi/3)$.
The transition appears on $\mu_{\mathrm{I}}/T=\pi/3$ at high temperature above $T_{\mathrm{RW}}$ where $T_{\mathrm{RW}}$ is the RW endpoint.
Using these properties, we proceed our calculation and analysis.
\\
\indent
In order to study the large real $\mu$ region, we use the two-phase model consisting of the quantum hadrodynamics (QHD) model for the hadron phase and the entanglement-PNJL (EPNJL) model for the quark phase.
The QHD model is satisfied with the nuclear saturation properties~\cite{Lalazissis} and the EPNJL model reproduces well the full LQCD results at zero and imaginary $\mu$~\cite{Sakai,Ishii}.
In addition, we determine the strength of the vector interaction for the EPNJL model using the quark number density calculated with LQCD.
Then, using the two-phase model, we discuss the mass-radius (MR) relation of the NSs and explore the hadron-quark phase transition line.
\section{Quark number density}
\label{QND}
The quark number density ($n$) is defined as
\beq
\frac{n}{T^{3}}=\frac{1}{VT^{2}}\frac{\partial}{\partial \mu}\ln Z = \frac{N_{f}N_{t}^{3}}{N_{V}}\mathrm{tr}\left[\Delta^{-1}\frac{\partial \Delta}{\partial \hat{\mu}} \right], \label{eq:nq}
\eeq
where $T$ is the temperature, $V$ is the volume, $Z$ is the QCD partition function, $N_{f}$ is the number of flavors, $N_{t}$ is the temporal lattice size, $N_{V}$ is the lattice volume, $\hat{\mu}$ is the chemical potential in lattice units and $\Delta$ is the fermion matrix.
We apply the random noise method for the trace in Eq.(\ref{eq:nq}).
\\
\indent
The quark number density is an odd and a smooth function of $\mu$ in the hadron phase because there is no RW transition, but not smooth at $\mu_{\mathrm{I}}/T=\pi/3$ in the quark phase because there exists the RW transition.
In order to extrapolate the quark number density at imaginary $\mu$, we fit them by using the Fourier series in the hadron phase,
\beq
f^{1}_{F}\left(\frac{\mu_{\mathrm{I}}}{T}\right) &=& F^{(1)}_{o} \sin\left(3 \frac{\mu_{\mathrm{I}}}{T}\right), \label{eq:f1F}\\
f^{2}_{F}\left(\frac{\mu_{\mathrm{I}}}{T}\right) &=& F^{(1)}_{o} \sin\left(3 \frac{\mu_{\mathrm{I}}}{T}\right) + F^{(2)}_{o} \sin\left(6 \frac{\mu_{\mathrm{I}}}{T}\right)\label{eq:f2F},
\eeq
and the polynomials of $\mu_{\mathrm{I}}/T$ in the quark phase,
\beq
f^{3}_{p}\left(\frac{\mu_{\mathrm{I}}}{T}\right) &=& p^{(1)}_{o} \left(\frac{\mu_{\mathrm{I}}}{T}\right) + p^{(3)}_{o} \left(\frac{\mu_{\mathrm{I}}}{T}\right)^{3}, \label{eq:f3p} \\
f^{5}_{p}\left(\frac{\mu_{\mathrm{I}}}{T}\right) &=& p^{(1)}_{o} \left(\frac{\mu_{\mathrm{I}}}{T}\right) + p^{(3)}_{o} \left(\frac{\mu_{\mathrm{I}}}{T}\right)^{3} + p^{(5)}_{o} \left(\frac{\mu_{\mathrm{I}}}{T}\right)^{5}\label{eq:f5p}.
\eeq
By replacing $\mu_{\mathrm{I}}/T$ by $\mu/T$, these are easily continued to real $\mu$:
\beq
f^{1}_{F}\left(\frac{\mu}{T}\right) &=& F^{(1)}_{o} \sinh\left(3 \frac{\mu}{T}\right), \label{eq:f1Fr} \\
f^{2}_{F}\left(\frac{\mu}{T}\right) &=& F^{(1)}_{o} \sinh\left(3 \frac{\mu}{T}\right) + F^{(2)}_{o} \sinh\left(6 \frac{\mu}{T}\right), \label{eq:f2Fr}\\
f^{3}_{p}\left(\frac{\mu}{T}\right) &=& p^{(1)}_{o} \left(\frac{\mu}{T}\right) - p^{(3)}_{o} \left(\frac{\mu}{T}\right)^{3}, \label{eq:f3pr}  \\
f^{5}_{p}\left(\frac{\mu}{T}\right) &=& p^{(1)}_{o} \left(\frac{\mu}{T}\right) - p^{(3)}_{o} \left(\frac{\mu}{T}\right)^{3} + p^{(5)}_{o} \left(\frac{\mu}{T}\right)^{5}\label{eq:f5pr}.
\eeq

\section{Results of the lattice simulations and their analytic continuation}
\begin{figure}[h]
\begin{center}
\hspace{-10pt}
 \includegraphics[angle=-90,width=0.445\textwidth]{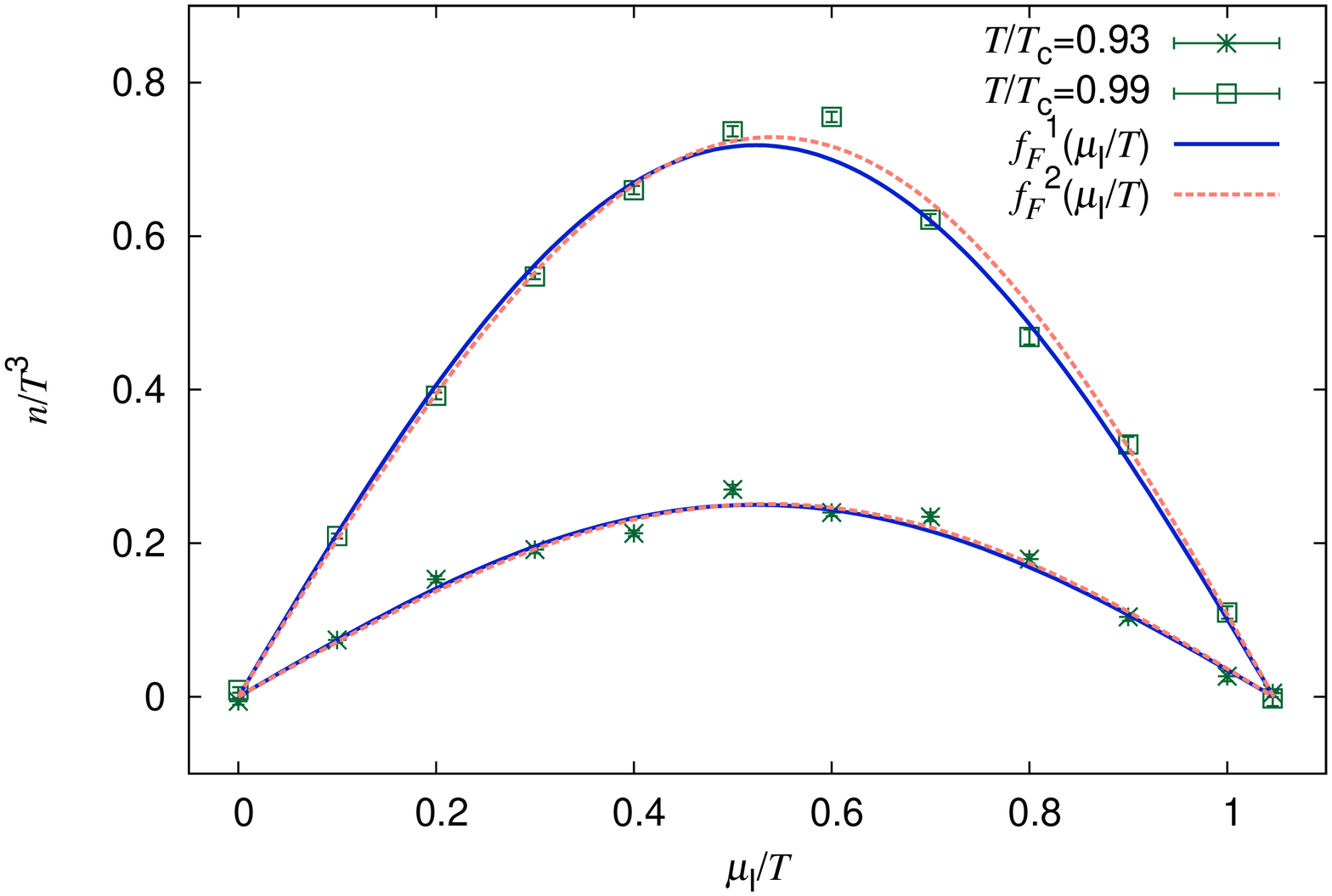}
 \includegraphics[angle=-90,width=0.445\textwidth]{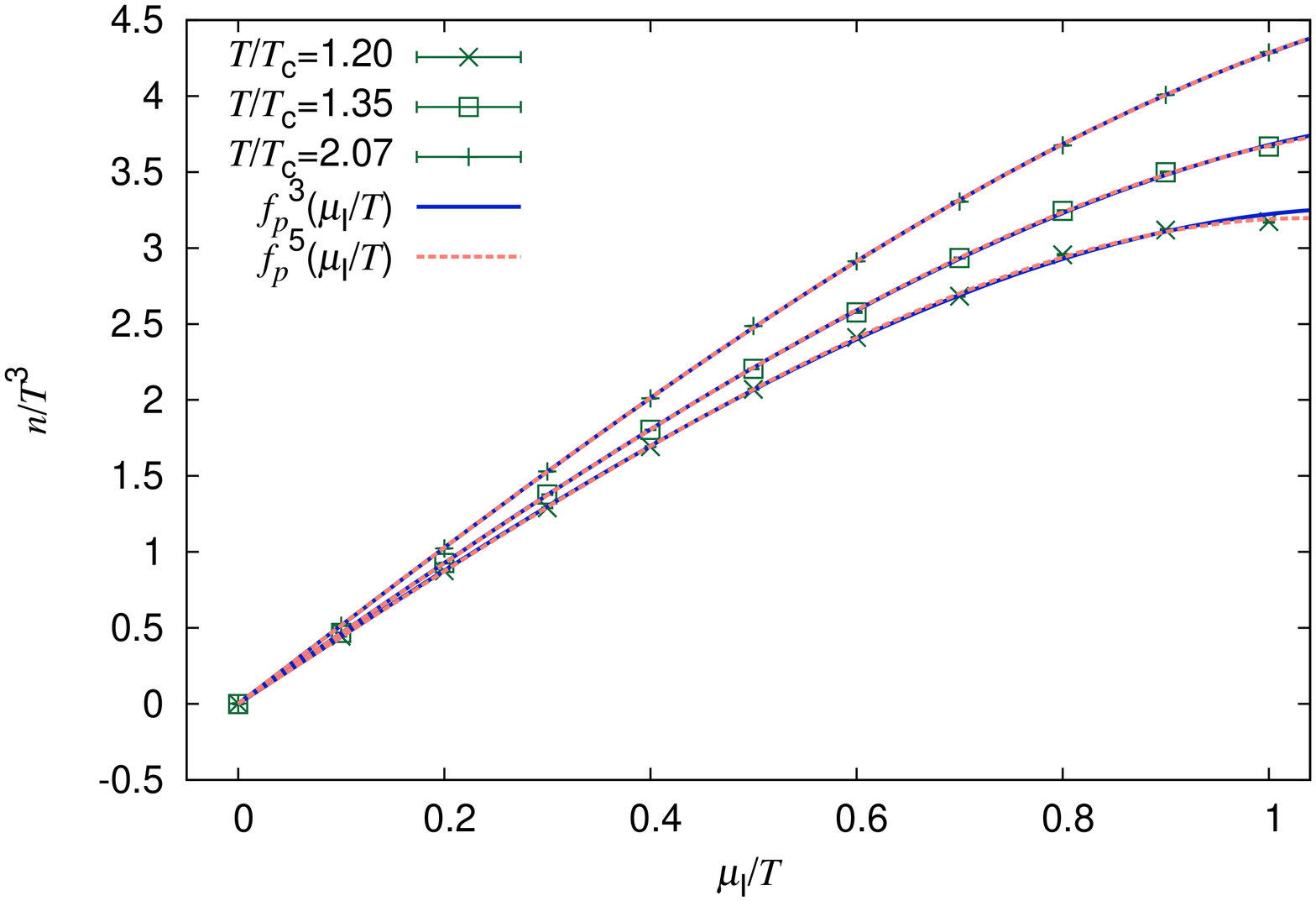}
\end{center}
\vspace{-10pt}
\caption{
The left panel shows $\mu_{\mathrm{I}}/T$ dependence of $n/T^3$ at $T< T_{\mathrm{c}}$.
Blue solid lines and red dashed lines are the results of fitting by $f^{1}_{F}(\mu_{\mathrm{I}}/T)$ and $f^{2}_{F}(\mu_{\mathrm{I}}/T)$, respectively.
The right panel shows $\mu_{\mathrm{I}}/T$ dependence of $n/T^3$ at $T> T_{\mathrm{RW}}$.
Blue solid lines and red dashed lines are the results of fitting by $f^{3}_{p}(\mu_{\mathrm{I}}/T)$ and $f^{5}_{p}(\mu_{\mathrm{I}}/T)$, respectively.
}
\label{mui-dep_nmbdst}
\end{figure}

\begin{figure}[h]
\begin{center}
\hspace{-10pt}
 \includegraphics[angle=-90,width=0.445\textwidth]{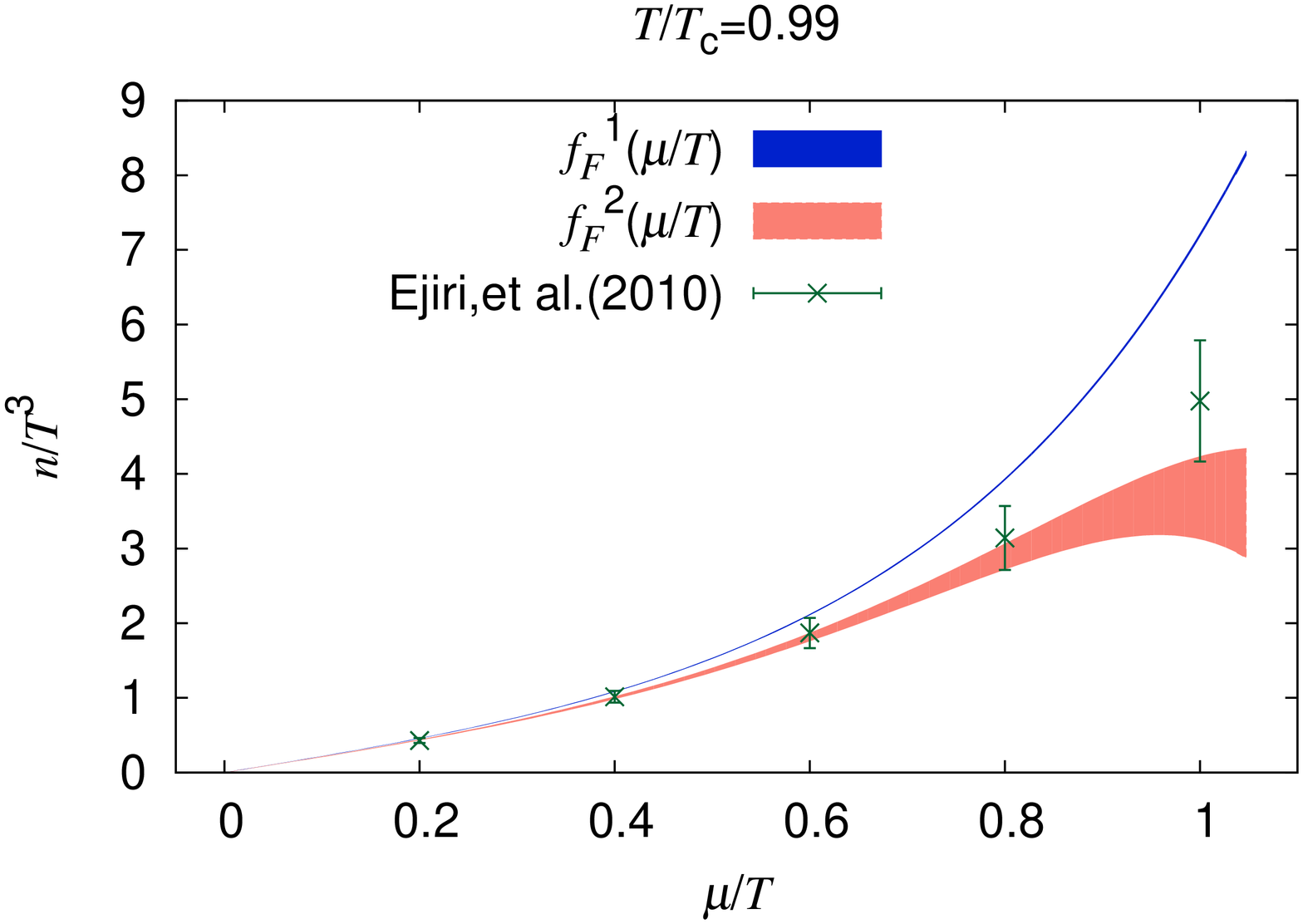}
 \includegraphics[angle=-90,width=0.445\textwidth]{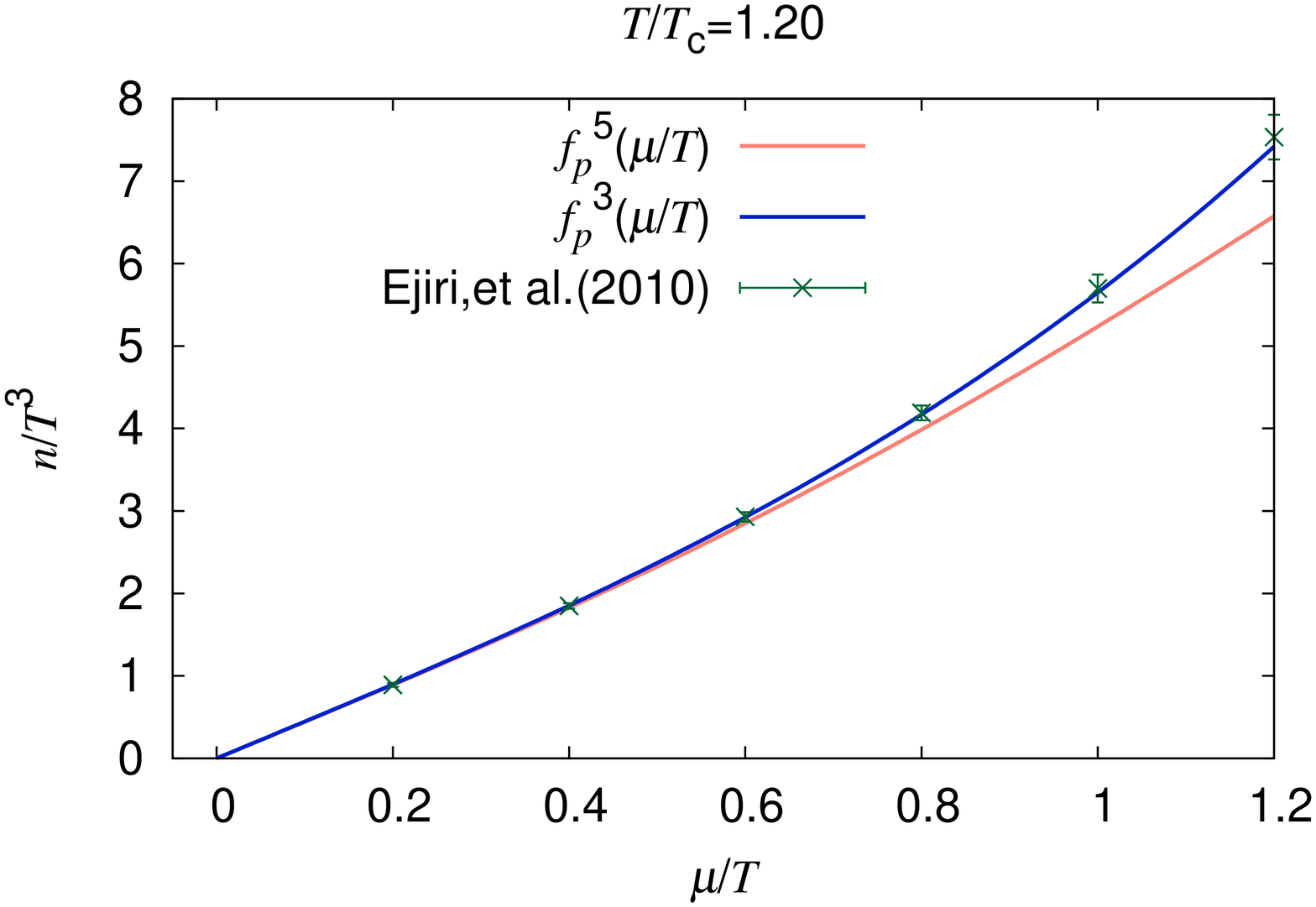}
\end{center}
\vspace{-10pt}
\caption{
The left panel shows $\mu/T$ dependence of $n/T^3$ at $T/T_{\mathrm{c}}=0.99$ by the analytic continuation.
Blue and red areas are the results of the analytic continuation by $f^{1}_{F}(\mu/T)$ and $f^{2}_{F}(\mu/T)$, respectively.
The right panel shows $\mu/T$ dependence of $n/T^3$ at $T/T_{\mathrm{c}}=1.20$ by the analytic continuation.
Blue and red lines are the results of the analytic continuation by $f^{3}_{p}(\mu/T)$ and $f^{5}_{p}(\mu/T)$, respectively.
The errors of the polynomial coefficients of $f^{3}_{p}(\mu/T)$ and $f^{5}_{p}(\mu/T)$ are within the thickness of lines.
}
\label{mur-dep_nmbdst}
\end{figure}

\begin{figure}[h]
\begin{center}
\hspace{-10pt}
 \includegraphics[angle=-90,width=0.445\textwidth]{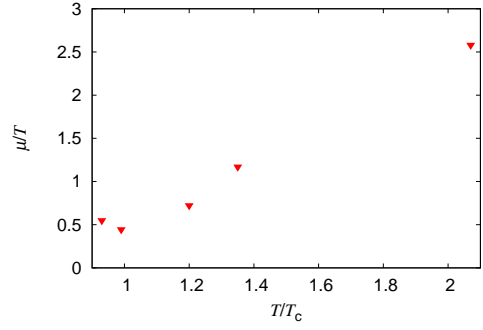}
\end{center}
\vspace{-10pt}
\caption{
The upper limit of the reliable extrapolated region as a function of temperature.
}
\label{xtrpl_T-mur3}
\end{figure}

We employed the clover-improved two-flavor Wilson fermion action and the renormalization-group improved Iwasaki gauge action.
The simulations were performed on a lattice of $N_{x} \times N_{y} \times N_{z} \times N_{t} = 8 \times 8 \times 16 \times 4$.
We computed the quark number densities along the line of constant physics with $m_{\mathrm{PS}}/m_{\mathrm{V}} = 0.80$~\cite{Maezawa}.
We considered five temperatures $T/T_{c} = 0.93$, $0.99$, $1.20$, $1.35$ and $2.07$ where $T_{c}$ is the critical temperature at $\mu=0$.
We measured the quark number densities at every 100 trajectories.
\\
\indent
Figure~\ref{mui-dep_nmbdst} shows $\mu_{\mathrm{I}}/T$ dependence of $n/T^{3}$ represented by the green symbols.
When the temperature is below $T_{c}$, $n/T^{3}$ behave as the sine function, and when the temperature is above $T_{\mathrm{RW}}$, $n/T^{3}$ increase monotonically up to $\mu_{\mathrm{I}}/T=\pi/3$.
Moreover, figure~\ref{mui-dep_nmbdst} shows our results of fitting our lattice results by eqs.(\ref{eq:f1F})-(\ref{eq:f5p}) in order to extrapolate the quark number density from imaginary to real $\mu$.
The two fittings give the same quality of agreement with lattice data for all temperatures.
\\
\indent
Figure~\ref{mur-dep_nmbdst} shows the quark number densities at real $\mu$ extrapolated from imaginary $\mu$ by eqs.(\ref{eq:f1Fr})-(\ref{eq:f5pr}) for $T/T_{c}=0.99$ on the left panel and for $T/T_{c}=1.20$ on the right panel.
In the previous study~\cite{Ejiri}, the Taylor expansion coefficients of the quark number densities up to 3rd order has been calculated at real chemical potential directly with the Taylor expansion method for the reweighting factors.
As one can see, at $T/T_{c}=1.20$, our result is consistent with the previous results.
\\
\indent
For $T<T_{c}$, as an estimate of the accuracy of the Fourier series up to the next to leading order, we assume that the Fourier series is reliable when the next to leading order contribution is smaller than 10 \% of the leading order contribution.
For $T>T_{\mathrm{RW}}$, as an estimate of the accuracy of the Taylor expansion series up to the 5th order, we assume that the expansion series is reliable when the 5th order contribution is smaller than 10 \% of the 3rd order contribution.
Figure~\ref{xtrpl_T-mur3} shows the upper limit of the reliable extrapolated region as a function of $T$.
The upper limit of the extrapolation from imaginary to real $\mu$ goes up as $T$ increases.
This indicate that the higher-order contributions become less important.
\section{Analyses by the effective model}
\begin{figure}[h]
\begin{center}
\hspace{-10pt}
 \includegraphics[angle=-90,width=0.445\textwidth]{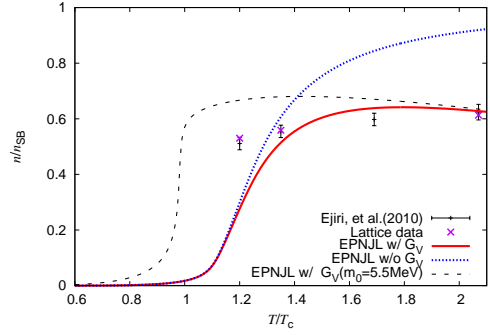}
\end{center}
\vspace{-10pt}
\caption{
The ratio $n/n_{\mathrm{SB}}$ as a function of temperature in the limit $\mu \rightarrow 0$.
}
\label{Tdep_n-n_SB}
\end{figure}

\begin{figure}[h]
\begin{center}
\hspace{-10pt}
 \includegraphics[angle=-90,width=0.445\textwidth]{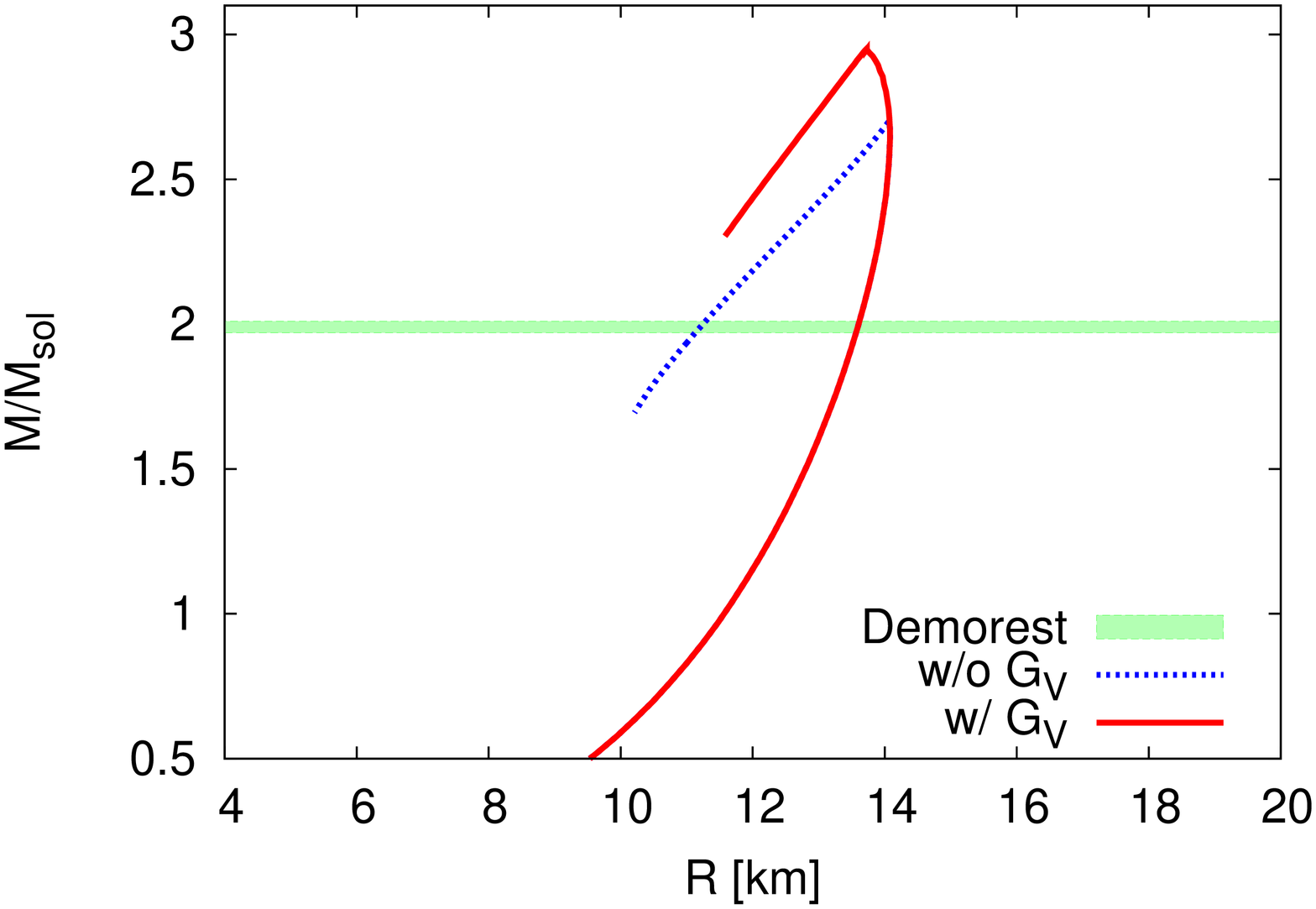}
 \includegraphics[angle=-90,width=0.445\textwidth]{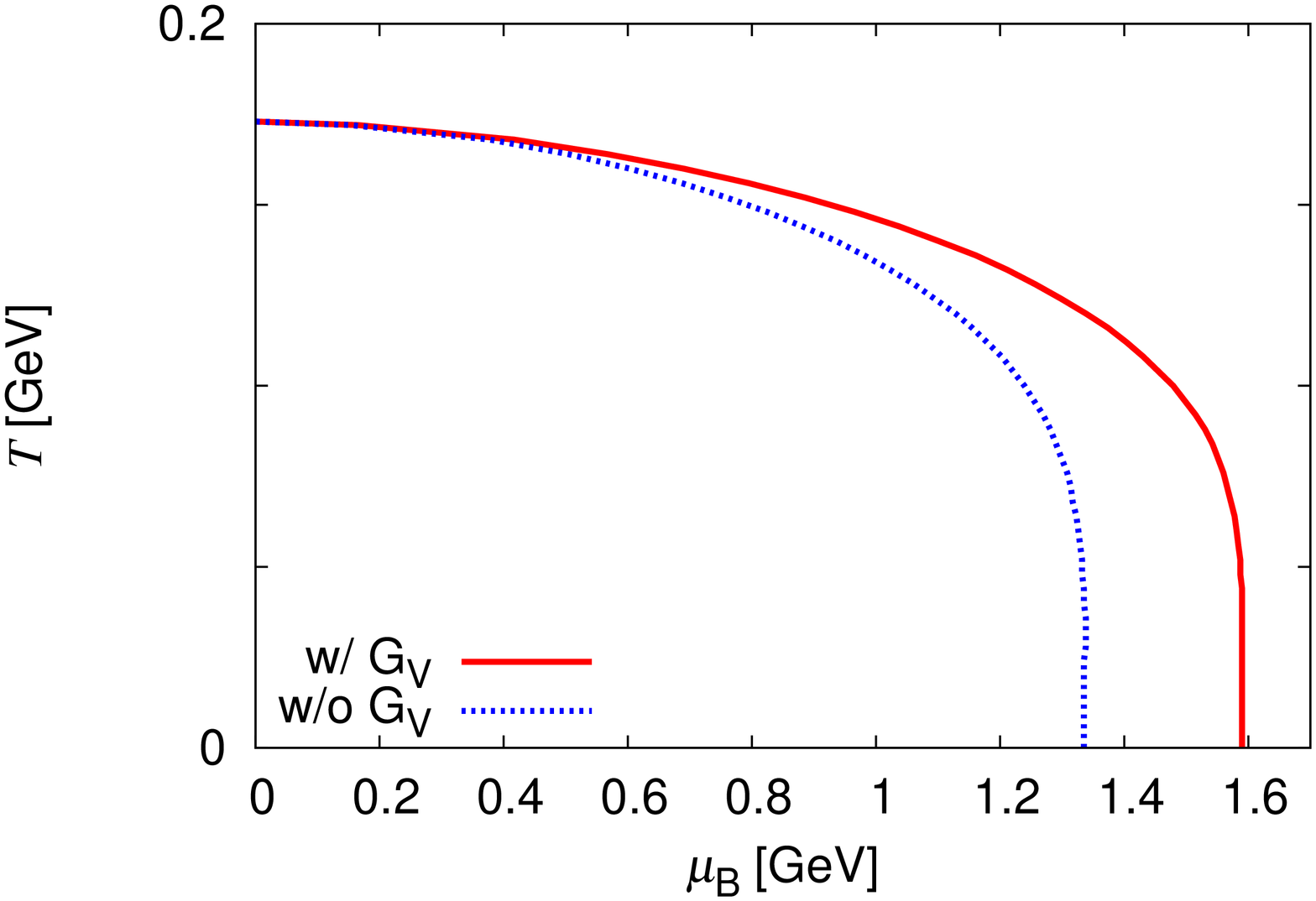}
\end{center}
\vspace{-10pt}
\caption{
The left panel shows the mass-radius relation of NSs.
The right panel shows the QCD phase diagram.
}
\label{MR_QCDphase}
\end{figure}

In order to extrapolate the quark number density to the large real $\mu$ region, we adopt the two-phase model with the mean field approximation.
This model consists of the EPNJL model for the quark phase and the QHD model for the hadron phase.
The phase is determined from the Gibbs condition.
\\
\indent
The Lagrangian density of the EPNJL model is
\beq
{\cal L}_{\mathrm{EPNJL}} &=&
\bar{q}(i\gamma^{\nu}D_{\nu}-m_{0})q
+G_{\mathrm{s}}(\Phi)\left[(\bar{q}q)^{2}+(\bar{q}i\gamma_{5}\vec{\tau}q)^{2}\right] \nonumber \\
&&-G_{\mathrm{v}}(\Phi)(\bar{q}\gamma_{\nu}q)^{2}
-{\cal U}(\Phi[A],\Phi^{\ast}[A],T),
\eeq
where $q$ is the quark field, $m_{0}$ is the current quark mass and $D^{\nu}=\partial^{\nu}+iA^{\nu}$ for $A^{\nu}=g\delta^{\nu}_{0}(A_{4})_{a}\lambda_{a}/2$, $(A_{\nu})_{a}$ is the gauge field, $\lambda_{a}$ is the Gell-Mann matrix and $g$ is the gauge coupling.
$G_{\mathrm{s}}(\Phi)$ and $G_{\mathrm{v}}(\Phi)$ are the coupling constants of the scalar- and vector-type four-quark interactions depending on the Polyakov loop $\Phi$,
\beq
G_{\mathrm{s}}(\Phi)=G_{\mathrm{s}}\left[1-\alpha_{1}\Phi \Phi^{\ast}-\alpha_{2}(\Phi^{3}+\Phi^{\ast 3})\right], \, \, \, G_{\mathrm{v}}(\Phi)=\alpha_{3}G_{\mathrm{s}}(\Phi).
\eeq
The effective potential ${\cal U}$ as a function of Polyakov loop is determined from lattice results in the pure gauge limit.
The parameters $\alpha_{1}=\alpha_{2}=0.2$, which are determined so as to reproduce well the full LQCD results for the deconfinement and chiral transition lines at zero and imaginary $\mu$~\cite{Sakai}.
The parameter $\alpha_{3}$ is determined from the full LQCD results for $n/n_{\mathrm{SB}}$ in the limit $\mu \rightarrow 0$.
\\
\indent
Figure~\ref{Tdep_n-n_SB} shows $T$ dependence of $n/n_{\mathrm{SB}}$ in the limit $\mu \rightarrow 0$.
In lattice calculations, $n$ is divided by the Stefan-Boltzmann (SB) limit for the lattice action in order to eliminate finite-volume effects. 
In model calculations, $n$ is divided by the SB limit in the continuum theory.
The cross symbols represent our lattice results.
These are consistent with the previous study~\cite{Ejiri}.
The blue-dotted and red-solid lines represent the results of the EPNJL model with $G_{v}=0$ and $G_{v}=0.33G_{s}$, respectively.
Obviously, the result of the EPNJL model with vector interaction is consistent with the LQCD results.
The dashed line represents the result of the EPNJL model with $G_{v}=0.33G_{s}$ in which $m_{0}$ set to the physical value $5.5$ MeV.
In ref.~\cite{Sugano}, it was shown that $m_{0}$ dependence of the ratio $\alpha_{3}=G_{v}/G_{s}$ is weak.
Therefore, in the following, we use the EPNJL model with $m_{0}=5.5$ MeV and $\alpha_{3}=0.33$ for the quark phase.
\\
\indent
The Lagrangian density of the QHD model is
\beq
{\cal L}_{\mathrm{QHD}} &=& \bar{\psi}(i\gamma^{\nu}\partial_{\nu}-m_{N}-g_{\sigma}\phi-g_{\omega}\gamma^{\nu}\omega_{\nu})\psi+\frac{1}{2}\partial^{\nu}\phi\partial_{\nu}\phi \nonumber \\
&&-\frac{1}{4}(\partial^{\mu}\omega^{\nu}-\partial^{\nu}\omega^{\mu})(\partial_{\mu}\omega_{\nu}-\partial_{\nu}\omega_{\mu})-U_{\mathrm{QHD}}, \\
U_{\mathrm{QHD}}&=&\frac{1}{2}m^{2}_{\sigma}\phi^{2}+\frac{1}{3}g_{2}\phi^{3}+\frac{1}{4}g_{3}\phi^{4}-\frac{1}{2}m^{2}_{\omega}\omega^{\nu}\omega_{\nu},
\eeq
where $\psi$, $\phi$, $\omega_{\nu}$, $m_{N}$, $m_{\sigma}$ and $m_{\omega}$ are nucleon($N$), $\sigma$-meson and $\omega$-meson fields and their masses, respectively, while $g_{\sigma}$, $g_{\omega}$, $g_{2}$ and $g_{3}$ are $\sigma$-$N$, $\omega$-$N$ and higher-order couplings, respectively.
We use the NL3 set~\cite{Lalazissis} as the parameter set of the QHD model.
\\
\indent
In Fig.~\ref{MR_QCDphase}, the left panel shows the MR relation of NSs where M$_{\mathrm{sol}}$ means the solar mass.
This relation is obtained by solving the Tolman-Oppenheimer-Volkoff equation.
Our results are satisfied with the observation data of the neutron star with twice a solar mass~\cite{Demorest}. 
The quark-hadron phase transition occurs where the MR-relation curve bends.
After this point, the neutron star has the quark phase in its inner core.
In Fig.~\ref{MR_QCDphase}, the right panel shows the phase diagram in the $\mu_{\mathrm{B}}$-$T$ plane for the hadron-quark phase transition.
$\mu_{\mathrm{B}}$ means the baryon chemical potential.
Our result of the two-phase model with $G_{v}=0.33G_{s}$ is $\mu_{c} \sim 1.6$ GeV which is the critical chemical potential at $T=0$.
This value is consistent with the previous study~\cite{Sasaki}.
When $G_{v}=0$, $\mu_{c}$ is shifted down.
Therefore, the contribution of the vector-type four-quark interactions is quite important.
\section{Summary}
We calculated the quark number density at imaginary $\mu$ by LQCD with clover-improved two-flavor Wilson fermion.
When the temperature is below $T_{c}$, the quark number densities behave as the sine function, and when the temperature is above $T_{\mathrm{RW}}$, these increase monotonically up to $\mu_{\mathrm{I}}/T=\pi/3$.
The quark number densities are extrapolated from imaginary to real $\mu$ with analytic continuation.
For $T<T_{c}$, we used the Fourier series and, for $T>T_{\mathrm{RW}}$, we used the Taylor expansion series of $\mu_{\mathrm{I}}/T$.
Our extrapolated result at $T/T_{c}=1.20$ is consistent with the previous study~\cite{Ejiri}.
Furthermore, by estimating the higher order contribution, we showed that the analytic continuation of the quark number density from imaginary to real $\mu$ may be valid up to $\mu/T \sim 0.8$.
\\
\indent
In order to study the large real $\mu$ region, we use the two-phase model consisting of the QHD model for the hadron phase and the EPNJL model for the quark phase.
First, we determine the strength of the vector-type four-quark interaction in the EPNJL model by using the results of the normalized quark number density $n/n_{\mathrm{SB}}$ calculated with LQCD.
The results indicate that, when the ratio $\alpha_{3}=G_{v}/G_{s}=0.33$, the EPNJL model reproduces well the LQCD data.
Next, we calculated the MR relation of NSs.
Our results are satisfied with the observation data of the neutron star with twice a solar mass~\cite{Demorest}. 
In addition, our present model predicts that the neutron star has the quark phase in its inner core.
Finally, we explored the hadron-quark phase transition in $\mu_{\mathrm{B}}$-$T$ plane.
Our result of the two-phase model with $G_{v}=0.33G_{s}$ is $\mu_{c} \sim 1.6$ GeV, which is consistent with the previous study~\cite{Sasaki}.
When $G_{v}=0$, $\mu_{c}$ is shifted down.
Thus, the contribution of the vector-type four-quark interactions is quite important.
\noindent
\begin{acknowledgments}
We thank A. Nakamura, K. Nagata and T. Sasaki for useful discussions.
M. Y., H. K., and J. T. are supported by Grant-in-Aid for Scientific Research (No. 26400278, No. 26400279, and No. 25-3944) from the Japan Society for the Promotion of Science (JSPS).
The numerical calculations were performed on NEC SX-9 and SX-8R at CMC, Osaka University and HITACHI HA8000 at Research Institute for Information Technology, Kyushu University.
\end{acknowledgments}

%
\end{document}